\documentclass[onecolumn]{revtex4}

\usepackage{graphicx}
\usepackage{dcolumn}
\usepackage{bm}

\input epsf

\begin{document}

\title{\Large Study of Generalized Second Law of Thermodynamics
in Loop Quantum Cosmology with the Effect of Non-Linear
Electrodynamics}

\author{\bf~Tanwi~Bandyopadhyay$^1$\footnote{tanwib@gmail.com}~and~Ujjal~Debnath$^2$
\footnote{ujjaldebnath@yahoo.com,ujjal@iucaa.ernet.in}}

\affiliation{$^1$Department of Mathematics,~Shri~Shikshayatan
College, 11, Lord~Sinha Road,~Kolkata-71, India.\\ $^2$Department
of Mathematics,~Bengal Engineering and Science
University,~Shibpur,~Howrah-711103, India.}

\begin{abstract}
In this work, we have discussed the Maxwell's electrodynamics in
non-linear forms in FRW universe. The energy density and pressure
for non-linear electrodynamics have been written in the
electro-magnetic universe. The Einstein's field equations for flat
FRW model in loop quantum cosmology have been considered if the
universe is filled with the matter and electro-magnetic field. We
separately assumed the magnetic universe and electric universe.
The interaction between matter and magnetic field have been
considered in one section and for some particular form of
interaction term, we have found the solutions of magnetic field
and the energy density of matter. We have also considered the
interaction between the matter and electric field and another form
of interaction term has been chosen to solve the field equations.
The validity of generalized second law of thermodynamics has been
investigated on apparent and event horizons using Gibb's law and
the first law of thermodynamics for magnetic and electric universe
separately.
\end{abstract}

\maketitle

\sloppy \tableofcontents

\section{\normalsize\bf{Introduction}}

Observations of the redshift of supernovae type Ia
\cite{Riess,Perlm} and cosmic microwave background
\cite{Bennet,Hal} show that our universe is expanding with
acceleration and lead to the search for a new type of matter which
violates the strong energy condition, known as dark energy. The
dark energy has the property that the energy component produce
sufficient negative pressure, which drives the cosmic
acceleration. There are many candidates supporting this behavior.
Cosmological constant $\Lambda$ is the most popular candidate of
dark energy satisfying the equation of state (EoS)
$p_{\Lambda}=-\rho_{\Lambda}$. There are other strong favored
candidates like quintessence, which is composed by a scalar field
\cite{Caldwell,Peebles} with self-interacting potential. The EoS
of the fluid distribution of the universe is given by
$p=\omega\rho$, where $\omega$ is known as the EoS parameter. The
universe will accelerate if $\omega<-1/3$. $\omega=-1$ represents
$\Lambda$CDM model and $\omega<-1$ corresponds to the phantom
dominated model \cite{Melch}. Recently, many candidates with
variable EoS play the crucial role of dark energy to drive the
acceleration of the universe namely, Chaplygin gas
\cite{Kamen,Debnath}, Tachyonic field \cite{Sen}, holographic dark
energy \cite{Wang}, agegraphic dark energy \cite{Wei}, Ricci dark
energy \cite{Gao}, Hessence \cite{Cai}, DBI-essence \cite{Martin},
K-essence \cite{Picon}, dilaton dark energy \cite{Lu} etc.\\

At present, we live in an epoch where the dark energy and the dark
matter are comparable. For this purpose, the interacting dark
energy models have been studied to explain the cosmic coincidence
problem \cite{Jamil}. Till now, many works have been proposed for
this interacting dark energy \cite{Hu,Wu}. In recent years, the
model of interacting dark energy has been explored in the
framework of loop quantum cosmology (LQC). The LQC is the
application in the cosmological context of loop quantum gravity
(LQG) \cite{Rovelli,Thiemann,Lowan}, which is a theory trying to
quantize the gravity with a non-perturbative and background
independent method. By studying the early universe inflation and
the fate of future singularity in LQC, it is found that the big
bang singularity, the big rip singularity and other future
singularities can be avoided \cite{Ash}. It has been verified
that, the cosmological evolution in LQC for quintessence model is
same as that in classical Einstein cosmology, whereas for the
phantom dark energy, the loop quantum effect significantly reduce
the parameter spacetime required by stability. Recently, the
dynamics of phantom, quintom and hessence dark energy models in
LQC have been studied \cite{Samart}.\\

On the other hand, a new approach \cite{De} has recently been
taken to avoid the cosmic singularity through a non-linear
extension of the Maxwell's electromagnetic theory. The associated
Lagrangian and the resulting electrodynamics can theoretically be
justified, based on different arguments. The homogeneous and
isotropic non-singular FRW solutions can be obtained \cite{De} by
considering a generalized model of Maxwell's electrodynamics,
where local covariant and gauge-invariant Lagrangian depend on the
field invariants up to the second order, as a source of classical
Einstein's equations. Exact solutions of the Einstein's field
equations coupled to non-linear electrodynamics (NLED) may hint at
the relevance of the non-linear effects in strong gravitational
and magnetic fields. An inhomogeneous and anisotropic non-singular
model for the early universe filled with Born-Infeld type
non-linear electromagnetic field was studied \cite{Garcia}.
Recently, there are several works on NLED in
various situations \cite{Salim,Munoz,Berg,Camara}.\\

In Einstein's gravity, the connection between black hole
thermodynamics and Einstein's equations was first discovered in
\cite{Ted} by deriving the Einstein's equations from the
proportionality of entropy and horizon area together with the
first law of thermodynamics. The thermodynamical laws also have
been applied in the cosmological context, considering universe as
a thermodynamical system bounded by the apparent horizon. At the
apparent horizon, the first law of thermodynamics is equivalent to
the Friedmann equations and generalized second law (GSL) is obeyed
at the horizon. There are several studies in thermodynamics for
dark energy filled universe on apparent and event horizons
\cite{GWW,Gong,Paddy,Cao,Setare}. Thermodynamical properties in
non-linear electrodynamics has extensively studied in
\cite{Gonzalez}. In our previous work \cite{Bandyo}, we have
assumed the apparent, event, Hubble and particle horizons of the
FRW universe and have studied the first law and GSL of
thermodynamics in non-linear electrodynamics with magnetic field
only. In the present work, we extend the previous work in LQC by
considering both the electric and magnetic fields and examine the
validity of GSL in apparent and event horizons of the magnetic
universe and electric universe.\\

The investigation has been done in the following way : We have
briefly discussed the Maxwell's electrodynamics in non-linear
forms in section II. The energy density and pressure for
non-linear electrodynamics have been written in the
electro-magnetic universe. The Einstein's field equations for flat
FRW model in loop quantum cosmology have been considered if the
universe is filled with the matter and electro-magnetic field. We
separately assumed the magnetic universe ($E=0$) and electric
universe ($B=0$). The interaction between matter and magnetic
field have been considered in section III and for some particular
form of interaction term, we have found the solutions of magnetic
field and the energy density of matter. In section IV, we have
considered the interaction between the matter and electric field
only and another form of interaction term has been chosen to solve
the field equations. In section V, the validity of generalized
second law of thermodynamics have been investigated on apparent
and event horizons using Gibbs' law and the first law of
thermodynamics for magnetic universe and electric universe
separately. Finally, some conclusions are drawn in section IV.

\section{\normalsize\bf{Basic Equations in Non-linear Electrodynamics}}

The Lagrangian density in Maxwell's electrodynamics (linear) can
be written as \cite{Camara}

\begin{equation}
{\cal
L}=-\frac{1}{4\mu_{0}}~F^{\mu\nu}F_{\mu\nu}=-\frac{1}{4\mu_{0}}~F
\end{equation}

where $F^{\mu\nu}$ is the electromagnetic field strength tensor
and $\mu_{0}$ is the magnetic permeability. The generalization of
Maxwell's electro-magnetic Lagrangian (non-linear) up to the
second order terms of the fields is given by \cite{Camara}

\begin{equation}
{\cal L}=-\frac{1}{4\mu_{0}}~F+\omega F^{2}+\eta F^{*2}
\end{equation}

where $\omega$ and $\eta$ are arbitrary constants and

\begin{equation}
F^{*}\equiv F^{*}_{\mu\nu}F^{\mu\nu}
\end{equation}

where $F^{*}_{\mu\nu}$ is the dual of $F_{\mu\nu}$. Here we
consider the FRW model of our universe. Since the spatial section
of FRW geometry are isotropic, electromagnetic field can generate
such a universe only if an averaging procedure is performed
\cite{Tolman, Hind}. In this situation, the energy density and the
pressure of the NLED field should be evaluated by averaging over
volume. So the corresponding energy-momentum tensor for non-linear
electro-magnetic theory has the form

\begin{equation}
T_{\mu\nu}=-4~\frac{\partial {\cal L}}{\partial
F}~F^{\alpha}_{\mu}F_{\alpha\nu}+\left(\frac{\partial {\cal
L}}{\partial F^{*}}~F^{*}-{\cal L}\right)g_{\mu\nu}
\end{equation}

The modified Lagrangian in non-linear electrodynamics for
accelerated universe is considered as \cite{Garc}

\begin{equation}
{\cal L}=-\frac{1}{4}~F+\alpha F^{2}+\beta F^{-1}
\end{equation}

where $\alpha$ and $\beta$ are arbitrary (constant) parameters. As
seen this Lagrangian contains both positive and negative powers of
$F$. The second (quadratic) term dominates during very early
epochs of the cosmic dynamics, while the Maxwell term (first term
above) dominates in the radiation era. The last term is
responsible for the accelerated phase of the cosmic evolution
\cite{Novello}. The above Lagrangian density yields a unified
scenario to describe both the acceleration of the universe (for
weak fields) and the avoidance of the initial singularity, as a
consequence of its properties in the strong-field regime.\\

The energy density and pressure for electro-magnetic (EM) field
are given by \cite{Garc}

\begin{equation}
\rho_{F}=-{\cal L}-4E^{2}{\cal L}_{F}
\end{equation}
and
\begin{equation}
p_{F}={\cal L}-\frac{4}{3}(2B^{2}-E^{2}){\cal L}_{F}
\end{equation}

where, $B$ and $E$ are respectively magnetic field and electric
field. Now, the electro-magnetic field has the expression
$F=2(B^{2}-E^{2})$, so the explicit forms of the energy density
and the pressure for electro-magnetic field will be \cite{Garc,
Sayani}

\begin{equation}
\rho_{F}=\frac{1}{2}(B^{2}+E^{2})-4\alpha(B^{2}-E^{2})(B^{2}+3E^{2})
-\frac{\beta(B^{2}-3E^{2})}{2(B^{2}-E^{2})^{2}}
\end{equation}

and

\begin{equation}
p_{F}=\frac{1}{6}(B^{2}+E^{2})-\frac{4\alpha}{3}(B^{2}-E^{2})(5B^{2}-E^{2})
+\frac{\beta(7B^{2}-5E^{2})}{6(B^{2}-E^{2})^{2}}
\end{equation}

Due to the loop quantum effect, the standard Friedmann equation in
LQC can be modified by adding a correction term as (for flat
model) \cite{Ashtekar0,Sami}

\begin{equation}
H^{2}=\frac{8\pi~G}{3}\rho_{total}\left(1-\frac{\rho_{total}}{\rho_{1}}\right)
\end{equation}

and

\begin{equation}
\dot{H}=-4\pi~G(\rho_{total}+p_{total})\left(1-\frac{2\rho_{total}}{\rho_{1}}\right)
\end{equation}

Also the energy-conservation equation is given by

\begin{equation}
\dot{\rho}_{total}+3H(\rho_{total}+p_{total})=0
\end{equation}

with, $\rho_{total}=\rho_{m}+\rho_{F}$ and $p_{total}=p_{m}+p_{F}$
where, $\rho_{m}$ and $p_{m}$ are energy density and pressure for
matter obeying the equation of state $p_{m}=w_{m}\rho_{m}$,
$\rho_{1}=\sqrt{3}\pi^{2}\gamma^{3}G^{2}\hbar$ is called the
critical loop quantum density, $\gamma$ is the dimensionless
Barbero-Immirzi parameter and $H=\frac{\dot{a}}{a}$ is the Hubble parameter.\\

Now if we consider the homogeneous electric field $E$ in plasma,
it gives rise to an electric current of charged particles and then
rapidly decays. So the squared magnetic field $B^{2}$ dominates
over $E^{2}$, i.e., in this case, the average value
$<E^{2}>\approx 0$ and hence $F=2B^{2}$. So $F$ is now only the
function of magnetic field (vanishing electric component) and
hence the FRW universe may be called the {\it magnetic universe}.
Similarly, if the average value $<B^{2}>\approx 0$, then
$F=-2E^{2}$. So $F$ is now only the function of electric field
(vanishing magnetic component) and hence the FRW universe may be
called the {\it electric universe}. In the following two sections,
we shall assume the magnetic and electric universes separately. \\

\section{\normalsize\bf{Interaction between matter and Magnetic field}}

In the magnetic universe, we assume $E=0$. Therefore the energy
density and pressure are (using equations (8) and (9))

\begin{equation}
\rho_{B}=\frac{1}{2}B^{2}-4\alpha B^{4}-\frac{\beta}{2B^{2}}
\end{equation}

\begin{equation}
p_{B}=\frac{1}{6}B^{2}-\frac{20\alpha}{3}B^{4}+\frac{7\beta}{6B^{2}}
\end{equation}

and subsequently

\begin{equation}
\rho_{total}=\rho_{m}+\frac{1}{2}B^{2}-4\alpha
B^{4}-\frac{\beta}{2B^{2}}
\end{equation}

and

\begin{equation}
p_{total}=\omega_{m}\rho_{m}+\frac{1}{6}B^{2}-\frac{20\alpha}{3}B^{4}+\frac{7\beta}{6B^{2}}
\end{equation}

Now if we consider the interaction between matter and magnetic
field, then the conservation equation (12) becomes

\begin{equation}
\dot{\rho}_{m}+3H(1+w_{m})\rho_{m}=Q
\end{equation}

and

\begin{equation}
\dot{\rho}_{B}+3H(\rho_{B}+p_{B})=-Q
\end{equation}

where $Q$ is the interaction term. Now let us take the interaction
component $Q$ as

\begin{equation}
Q=2\delta B\left(1-16\alpha B^{2}+\frac{\beta}{B^{4}}\right)H
\end{equation}

This expression is used to simplify the calculation procedure.
Here $\delta$ may be treated as interaction parameter which is a
small positive quantity. Using the expressions of $\rho_{B}$,
$p_{B}$ from (15) and (16) and the relation between $\rho_{m}$ and
$p_{m}$, the above two equations (17) and (18) give the solutions
of $B$ and $\rho_{m}$ in the following forms:

\begin{equation}
B=-\delta+\frac{B_{0}}{a^{2}}
\end{equation}

and

\begin{eqnarray*}
\rho_{m}=\rho_{0}a^{-3(1+\omega_{m})}+\frac{2\delta}{a^{6}}\left[-\frac{a^{6}(\beta+\delta^{4}
-16\alpha\delta^{6})}{3\delta^{3}(1+\omega_{m})}+\frac{16\alpha
B_{0}^{3}}{3(1-\omega_{m})}+\frac{48a^{2}\alpha\delta
B_{0}^{2}}{3\omega_{m}-1}\right]
\end{eqnarray*}
\begin{eqnarray*}
+\frac{2B_{0}}{a^{2}\delta^{3}(1+3\omega_{m})}
\left(-3\beta+\delta^{4}-48\alpha\delta^{6}+6\beta~
_{2}F_{1}[\frac{1}{2}(1+3\omega_{m}),1,\frac{3(1+\omega_{m})}{2},\frac{a^{2}\delta}{B_{0}}]\right.
\end{eqnarray*}
\begin{equation}
\left.-4\beta ~
_{2}F_{1}[\frac{1}{2}(1+3\omega_{m}),2,\frac{3(1+\omega_{m})}{2},\frac{a^{2}\delta}{B_{0}}]
+\beta ~
_{2}F_{1}[\frac{1}{2}(1+3\omega_{m}),3,\frac{3(1+\omega_{m})}{2},\frac{a^{2}\delta}{B_{0}}]
\right)
\end{equation}

where $B_{0}$ and $\rho_{0}$ are positive integration constants and $_{2}F_{1}$ is the hypergeometric function.\\

\section{\normalsize\bf{Interaction between matter and Electric field}}

This section deals with the case $B=0$ i.e., the universe is
filled with matter and electric field. Thus the expressions of the
energy density and pressure for the electric field become (using
equations (8) and (9))

\begin{equation}
\rho_{E}=\frac{1}{2}E^{2}+12\alpha E^{4}+\frac{3\beta}{2E^{2}}
\end{equation}

\begin{equation}
p_{E}=\frac{1}{6}E^{2}-\frac{4\alpha}{3}E^{4}-\frac{5\beta}{6E^{4}}
\end{equation}

and therefore

\begin{equation}
\rho_{total}=\rho_{m}+\frac{1}{2}E^{2}+12\alpha
E^{4}+\frac{3\beta}{3E^{2}}
\end{equation}

and

\begin{equation}
p_{total}=\omega_{m}\rho_{m}+\frac{1}{6}E^{2}-\frac{4\alpha}{3}E^{4}-\frac{5\beta}{6E^{2}}
\end{equation}

In this case, we consider the interaction between matter and
electric field. So the conservation equation (12) takes the form

\begin{equation}
\dot{\rho}_{m}+3H(1+w_{m})\rho_{m}=\tilde{Q}
\end{equation}

and

\begin{equation}
\dot{\rho}_{E}+3H(\rho_{E}+p_{E})=-\tilde{Q}
\end{equation}

To simplify of the calculation, the interaction component
$\tilde{Q}$ is taken as

\begin{equation}
\tilde{Q}=3\tilde{\delta} E^{2}\left(1+16\alpha
E^{2}+\frac{\beta}{E^{4}}\right)H
\end{equation}

where $\tilde{\delta}$ is a small positive quantity. In this case,
exact solutions for $\rho_{m}$ and $E$ from the above two
equations can only be obtained if they are solved numerically
using the expressions of $\rho_{E}$ and $p_{E}$
together with the relation between $\rho_{m}$ and $p_{m}$.\\

\section{\normalsize\bf{Generalized Second Law of Thermodynamics}}

In this section, the validity of the generalized second law of
thermodynamics is studied. It states that, the sum of entropy of
total matter enclosed by the horizon and the entropy of the
horizon does not decrease with time. In the following, we consider
apparent and event horizons only. The variation of entropy inside
the horizon will be calculated via Gibb's equation and the
variation of entropy on the horizon will be calculated using first
law of thermodynamics. Hence we shall examine the validity of GSL
of thermodynamics of the magnetic universe and electric universe
separately bounded by the above mentioned horizons.\\

\subsection{\normalsize\bf{Apparent Horizon}}

We know that radius of apparent horizon in the FRW universe,

\begin{equation}
R_{A}=\frac{1}{H}
\end{equation}

Therefore,

\begin{equation}
\dot{R}_{A}=-\frac{\dot{H}}{H^{2}}
\end{equation}

{\bf Case-I: $E=0$:} Using equations (10), (11), (15) and (16)

\begin{equation}
\dot{R}_{A}=\frac{B^{2}\left(1-16\alpha
B^{2}+\frac{\beta}{B^{4}}\right)+\frac{3}{2}(1+w_{m})\rho_{m}}{\rho_{m}+\frac{B^{2}}{2}-4\alpha
B^{4}-\frac{\beta}{2B^{2}}}
\end{equation}

Considering the net amount of energy crossing through the apparent
horizon in time $dt$ as \cite{Bousso}

\begin{equation}
-dE=4\pi R_{A}^{3}H(\rho_{total}+p_{total})dt
\end{equation}

and assuming the validity of first law of thermodynamics on the
apparent horizon, i.e.,

\begin{equation}
-dE=T_{A}dS_{A}
\end{equation}

where, $S_{A}$  and  $T_{A}$ are the entropy and temperature on
the apparent horizon, we have

\begin{equation}
\frac{dS_{A}}{dt}=\frac{4\pi
R_{A}^{3}H}{T_{A}}\left[\frac{2B^{2}}{3}\left(1-16\alpha
B^{2}+\frac{\beta}{B^{4}}\right)+(1+w_{m})\rho_{m}\right]
\end{equation}

Again from the Gibb's eqn

\begin{equation}
T_{A}dS_{I}=dE_{I}+p_{total}dV
\end{equation}

where, $S_{I}$ is the internal entropy, $V$ is the volume and
$E_{I}=\rho V$ is the internal energy, we have

\begin{equation}
\frac{dS_{I}}{dt}=\frac{4\pi
R_{A}^{2}}{T_{A}}\left[\frac{2B^{2}}{3}\left(1-16\alpha
B^{2}+\frac{\beta}{B^{4}}\right)+(1+w_{m})\rho_{m}\right](\dot{R}_{A}-HR_{A})
\end{equation}

From eqns (34) and (36), the rate of change of the total entropy
becomes

\begin{equation}
\frac{d}{dt}(S_{A}+S_{I})=\frac{4\pi
R_{A}^{2}}{T_{A}}\left[\frac{2B^{2}}{3}\left(1-16\alpha
B^{2}+\frac{\beta}{B^{4}}\right)+(1+w_{m})\rho_{m}\right]\dot{R}_{A}
\end{equation}

Substituting the expression (31) in eqn (37) and using the
expressions of $B$ and $\rho_{m}$ from eqns (20) and (21)
respectively, we plot the rate of change of total entropy of the
apparent horizon, i.e, $\dot{S}_{A}+\dot{S}_{I}$ against the scale
factor in figure 1 with interaction ($\delta=0.0001$) for
different matter components i.e., $w_{m}=1/3$ (solid line),
$w_{m}=0$ (dotted line) and $w_{m}=-0.5$ (dashed line). From the
figure, we see that the rate of change of total entropy for
apparent horizon is initially positive but as $a$ increases, it
decreases and in late time, it becomes negative. So we may
conclude that, under the loop quantum cosmological effects, the
GSL is valid initially, but after certain stage of the evolution
of the universe, the GSL will not valid for apparent horizon for
interacting scenarios in the magnetic universe.\\

\begin{figure}
\includegraphics[height=2.5in]{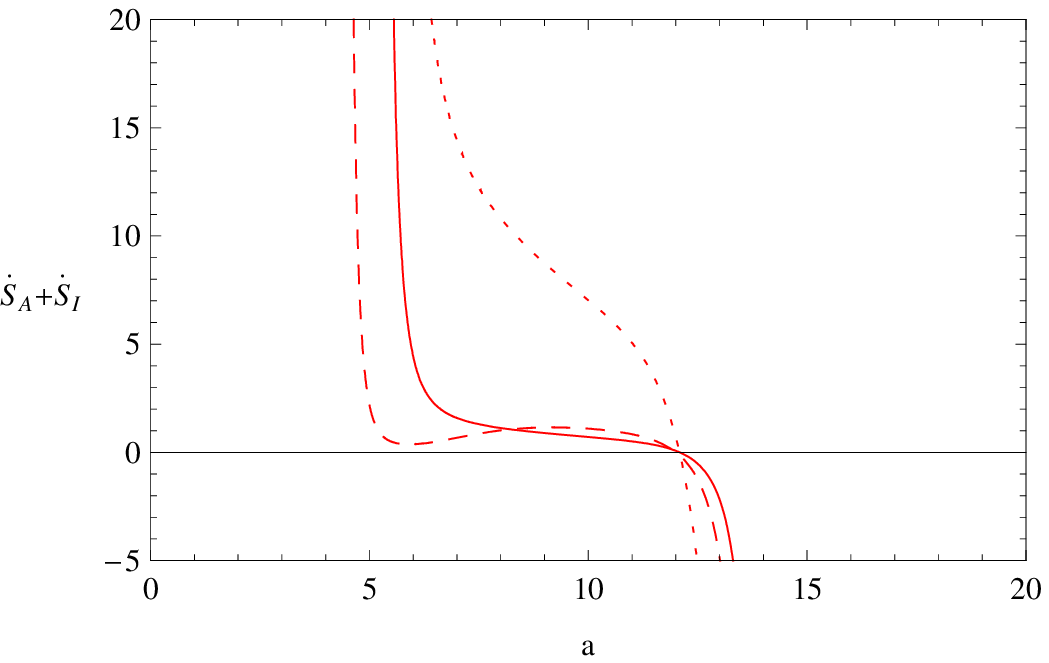}

\vspace{6mm} ~~~~~~~~~~~~Fig.1 represents rate of change of total
entropy of apparent horizon i.e., $\dot{S}_{A}+\dot{S}_{I}$
against the scale factor $a$ for $B=0$ with interaction for
$w_{m}=1/3$ (solid line), $w_{m}=0$ (dotted line) and $w_{m}=-0.5$
(dashed line).\\

\vspace{6mm}

\end{figure}

{\bf Case-II : $B=0$:}\\

Proceeding in the same way as in the previous case, the expression
for the rate of change of the total entropy becomes

\begin{equation}
\frac{d}{dt}(S_{A}+S_{I})=\frac{4\pi
R_{A}^{2}}{T_{A}}\left[\frac{2E^{2}}{3}\left(1+16\alpha
E^{2}+\frac{\beta}{E^{4}}\right)+(1+w_{m})\rho_{m}\right]\dot{R}_{A}
\end{equation}

Substituting the expression of $\dot{R}_{A}$ from eqn (30) and
using the numerical solutions of $E$ and $\rho_{m}$, we plot the
rate of change of total entropy of the apparent horizon, i.e,
$\dot{S}_{A}+\dot{S}_{I}$ against the scale factor in figure 2
with interaction ($\delta=0.0001$) for different matter components
i.e., $w_{m}=1/3$ (solid line), $w_{m}=0$ (dotted line) and
$w_{m}=-0.5$ (dashed line). From the figure, we see that the rate
of change of total entropy for apparent horizon is always
positive.Thus it may be concluded that even under the loop quantum
cosmological effects, the GSL is always satisfied in late time for
apparent horizon for interacting scenarios in the electric universe.\\

\begin{figure}
\includegraphics[height=2.5in]{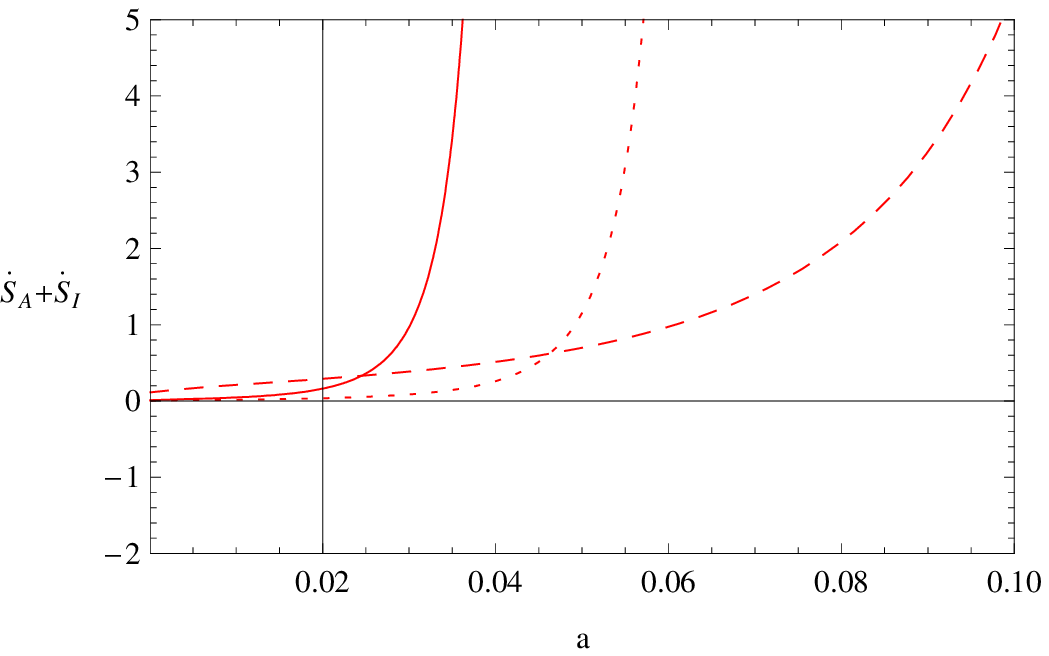}

\vspace{6mm} ~~~~~~~~~~~~Fig.2 represents rate of change of total
entropy of apparent horizon i.e., $\dot{S}_{A}+\dot{S}_{I}$
against the scale factor $a$ for $E=0$ with interaction for
$w_{m}=1/3$ (solid line), $w_{m}=0$ (dotted line) and $w_{m}=-0.5$
(dashed line).\\

\vspace{6mm}

\end{figure}

\subsection{\normalsize\bf{Event Horizon}}

The event horizon radius is given by

\begin{equation}
R_{E}=a\int_{a}^{\infty}\frac{da}{Ha^{2}}
\end{equation}

The differential eqn of which can be written as

\begin{equation}
\dot{R}_{E}=HR_{E}-1
\end{equation}

Considering the net amount of energy crossing through the event
horizon in time $dt$ as

\begin{equation}
-dE=4\pi R_{E}^{3}H(\rho_{total}+p_{total})dt
\end{equation}

and assuming the validity of first law of thermodynamics on the
event horizon, i.e,

\begin{equation}
-dE=T_{E}dS_{E}
\end{equation}

we have the rate of change of the total entropy as in the following cases:\\\\

{\bf Case-I : $E=0$:}

\begin{equation}
\frac{d}{dt}(S_{E}+S_{I})=\frac{4\pi
R_{E}^{2}}{T_{E}}\left[\frac{2B^{2}}{3}\left(1-16\alpha
B^{2}+\frac{\beta}{B^{4}}\right)+(1+w_{m})\rho_{m}\right](HR_{E}-1)
\end{equation}
\\

{\bf Case-II : $B=0$:}

\begin{equation}
\frac{d}{dt}(S_{E}+S_{I})=\frac{4\pi
R_{E}^{2}}{T_{E}}\left[\frac{2E^{2}}{3}\left(1+16\alpha
E^{2}+\frac{\beta}{E^{4}}\right)+(1+w_{m})\rho_{m}\right](HR_{E}-1)
\end{equation}

Substituting the expressions of ${R_{E}}$, $H$, $B$, $E$ and
$\rho_{m}$ in eqns (43) and (44), the rate of change of total
entropy of the event horizon, i.e, $\dot{S}_{E}+\dot{S}_{I}$ is
plotted against the scale factor in figures 3 and 4 for the above
two cases with interaction ($\delta=0.0001$) for different matter
components i.e., $w_{m}=1/3$ (solid line), $w_{m}=0$ (dotted line)
and $w_{m}=-0.5$ (dashed line). From the figures, we see that the
rate of change of total entropy for event horizon is always
positive for both magnetic and electric universes and hence the
GSL is always satisfied for event horizon for interacting
scenarios of the magnetic and electric universes.\\

\begin{figure}
\includegraphics[height=2.5in]{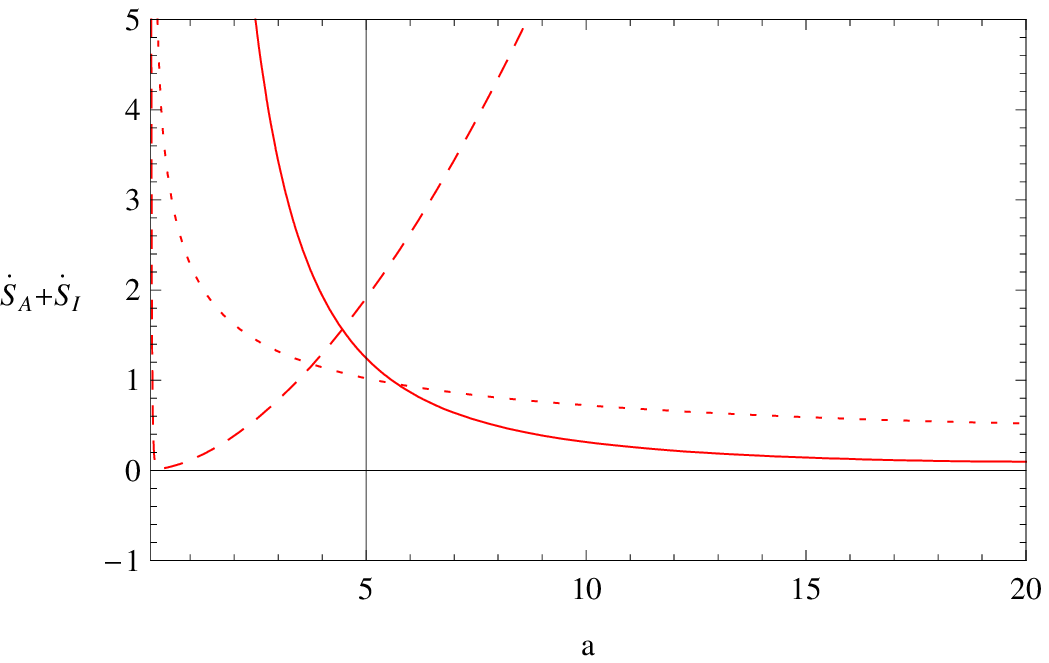}

\vspace{6mm} ~~~~~~~~~~~~{\bf Case-I : E=0:}~~~Fig.3 represents
rate of change of total entropy of event horizon i.e.,
$\dot{S}_{E}+\dot{S}_{I}$ against the scale factor $a$ for $B=0$
with interaction for $w_{m}=1/3$ (solid line), $w_{m}=0$ (dotted
line) and $w_{m}=-0.5$ (dashed line).\\

\vspace{6mm}

\end{figure}

\begin{figure}
\includegraphics[height=2.5in]{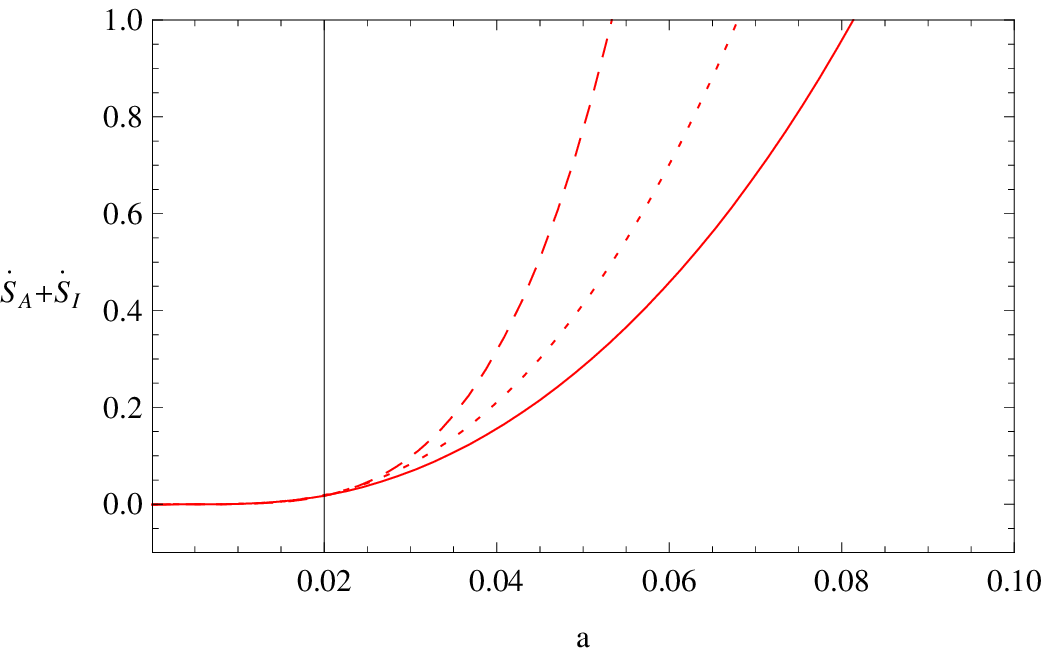}

\vspace{6mm} ~~~~~~~~~~~~{\bf Case-II : B=0}~~~Fig.4 represents
rate of change of total entropy of event horizon i.e.,
$\dot{S}_{E}+\dot{S}_{I}$ against the scale factor $a$ for $E=0$
with interaction for $w_{m}=1/3$ (solid line), $w_{m}=0$ (dotted
line) and $w_{m}=-0.5$ (dashed line).\\

\vspace{6mm}

\end{figure}

\section{\normalsize\bf{Discussions}}

In this work, we have briefly discussed Maxwell's electrodynamics
in non-linear forms for accelerating universe. The energy density
and pressure for non-linear electrodynamics have been written in
magnetic universe in one section and for universe with electric
field only in another section. The Einstein's field equations for
loop quantum cosmological model have been considered for FRW model
of the universe. The interaction between matter and electric and
magnetic fields have been incorporated separately and for
particular forms of interaction terms, we have found the solutions
for both electric as well as magnetic fields
and the energy density of matter.\\

In addition to this, our endeavor was to investigate the validity
of the generalized second law of thermodynamics of the universe
bounded by the apparent and event horizons. The variation of
entropy has been calculated inside the horizon using Gibb's
equation and that on the horizon using the first law of
thermodynamics. After that, we have studied the GSL of the universe
bounded by the above mentioned horizons.\\

In figures 1 - 4, the variation of total entropy on the apparent
and event horizons have been drawn against the scale factor $a$
for interacting scenarios ($\delta=0.0001$) of magnetic universe
as well as universe with electric field only, for
$w_{m}=0,1/3,-0.5$. From figure 1, we see that the rate of change
of total entropy was initially positive but in late epoch, it
becomes negative and thus the GSL is not valid for the magnetic
universe on the apparent horizon. Whereas on the event horizon for
the magnetic universe, the GSL remains always valid. This can be
clearly seen from figure 3. Figure 2 represents the rate of change
of total entropy on the apparent horizon for the universe with
electric field only, which again shows the validity of the GSL
throughout the evolution of the universe for all the cases.
Finally, from figure 4, we see that GSL was initially not valid on
the event horizon, but in late time it is satisfied for all types
of matter.\\

{\bf Acknowledgement}:\\

One of the authors (TB) wants to thank UGC, Govt. of India for
providing with a research project No. F.PSW-063/10-11 (ERO).\\\\

\end{document}